\documentstyle[aps,preprint,tighten,floats,epsf,rotate]{revtex}

\def\be{\begin{equation}}
\def\ee{\end{equation}}

\newcommand{\bpi}{\mbox{\boldmath $\pi$}}
\newcommand{\bx}{\mbox{\boldmath $x$}}
\newcommand{\br}{\mbox{\boldmath $r$}}
\newcommand{\bphi}{\mbox{\boldmath $\phi$}}
\newcommand{\btau}{\mbox{\boldmath $\tau$}}

\begin{document}
\draft
\preprint{\vbox{\it \null\hfill\rm    SI-TH-97-6, hep-ph/9712431}\\\\}
%
\title{Chiral phase transition
and baryon number conservation}
\author{G. Holzwarth \thanks{e-mail: holzwarth@hrz.uni-siegen.d400.de}}
\address{Fachbereich Physik, Universit\"{a}t-GH-Siegen, 
D-57068 Siegen, Germany} 
\date{June 1997}
\maketitle
\begin{abstract}
In the standard $R^4$ embedding of the chiral $O(4)$ model in 
3+1 dimensions the winding number is not conserved near 
the chiral phase transition and thus no longer can be identified with
baryon number. In order to reestablish conserved baryon 
number in effective low-energy models near and above the critical
temperature $T_c$ it is argued 
that insisting in $O(N)$ models on the angular nature of
the chiral fields with fixed boundary conditions restores 
conservation of winding number. For $N=2$ in 1+1 dimensions it is
illustrated that as a consequence of the angular boundary conditions
nontrivial solutions exist which would be unstable in $R^2$;
moving trajectories avoid crossing the origin; and time
evolution of random configurations after a quench leads to quasistable
soliton-antisoliton ensembles with net winding number fixed.
\end{abstract}
%

\vspace{1 cm}
\section{Introduction}
Effective chiral field theories in terms of a unitary
matrix field $U(\bx,t)\in SU(N_f)$  are considered as 
appropriate tool for the study of hadronic physics at very low 
energies~\cite{ChPT}. The chiral field $U$ comprises the lowest
pseudoscalar meson multiplet for $N_f$ flavors as Goldstone
bosons of the spontaneously broken chiral symmetry of QCD.
A most attractive feature of these models is the possibility to 
identify baryon number $B$ with the winding number~\cite{Skyrme} 
\be
\label{B}
B=\frac{1}{24 \pi^2} \int \;\epsilon^{i j k} {\rm tr} L_i L_j L_k
\;d^3x, ~~~~~~~L_i = U^\dagger \partial_i U,
\ee
that characterizes the map of compactified coordinate space onto 
the $SU(N_f)$ manifold of the chiral fields. This concept has a
profound basis~\cite{Wi83} in the anomaly structure 
of underlying QCD, and it has found
many successful applications as efficient description for
meson-baryon systems without explicit use of fermion fields.
Specifically, for $N_f=2$ flavors, it is the nontriviality of the
third homotopy group $\pi_3(SU(2))$ which allows for the existence 
of baryons as topological defects in the chiral pion field.

It is expected that with increasing temperature the spontaneously
broken chiral symmetry gets restored, such that beyond a critical
temperature the chiral condensate vanishes and the Goldstone bosons
acquire dynamical mass. 
Numerous examples in condensed matter systems~\cite{topo} show 
the decisive role of topological defects  
for the dynamics of phase transitions if spatial and internal
dimensions allow for nontrivial homotopy groups. 
For the cooling process of hot hadronic matter the topological
arguments have been used \cite{EKK} to obtain estimates for
the baryon-antibaryon yield
along the lines of the Kibble~\cite{Kib} mechanism.

On the other hand, it is expected that near and above the critical 
temperature $T=T_c$ it may be important to allow for additional
degrees of freedom. Apart from vector mesons it appears natural
to include (for $N_f=2$) the scalar partner of the pseudoscalar 
pions into a common chiral field
$\Phi=(\sigma,\bpi)$, and to relax the constraint to the $SU(2)$
3-sphere $\sigma^2+\bpi^2=f_\pi^2$. 
Thus the $SU(2)\times SU(2)$-symmetric nonlinear $\sigma$ model is
replaced by the $O(4)$-symmetric linear $\sigma$-model~\cite{GML}, 
where the fields 
$(\sigma,\bpi)$ can freely explore the full 4-dimensional chiral space.
This allows for a very convenient description of chiral symmetry 
restoration at finite temperatures~\cite{Baym},
to study critical fluctuations of the 
chiral condensate near the phase transition, or to follow the dynamical
formation of the condensate after a quench. 
It has been suggested that the exponential growth of collective
amplitudes in the dynamical evolution of chiral field configurations
after a quench may lead to disoriented chiral condensates in
macroscopic regions of space~\cite{ans} -\cite{LDC}. 
These considerations have been based~\cite{raj} -\cite{LDC} 
on calculations within the framework of the $O(4)$ model~\cite{GML}
with chiral fields embedded in the simply connected 
$R^4$ internal space.

Unfortunately, with the unconstrained embedding of the chiral field $\Phi$
into a simple $R^4$ manifold the concept of baryon number $B$ as
winding index of topological defects is lost because the 
topological connectedness of this manifold is trivial. 
Of course, also in this embedding we may
find nontrivial defect solutions which correspond to local minima in
the energy hypersurface. Although they may be separated from the vacuum 
configuration $\sigma \equiv f_\pi, \bpi \equiv 0 $ by a potential
barrier which prevents their decay within classical dynamics,
they no longer are topologically protected against
decay into mesonic fluctuations.
Cohen~\cite{Cohen} has argued that the corresponding quantum 
mechanical tunneling amplitude vanishes
in tree approximation in the limit where the number of colors 
$N_c \to \infty$. But a possible suppression 
for finite $N_c$ through quantum corrections to our knowledge
has never been proven.
By suitable changes in the parameters of the model (e.g. by increasing the
temperature) the decay process becomes even classically allowed.

Of course, we cannot have baryon number violated just by 
increasing the temperature. Apparently, we have to conclude that
within the linear $\sigma$-model it makes no sense to identify
winding number with baryon number. That conclusion, however, is quite
unsatisfactory: it would imply that we have lost a fundamental
symmetry of QCD in a model which we consider the appropriate tool
for the study of hadronic physics near $T_c$. We would naturally expect
that in cooling a hot hadronic plasma baryon-antibaryon production rates
depend crucially on respecting this symmetry.
\begin{figure}[h]
\begin{center}
\leavevmode
\vbox{\epsfxsize=14truecm \epsfysize=6truecm \epsfbox{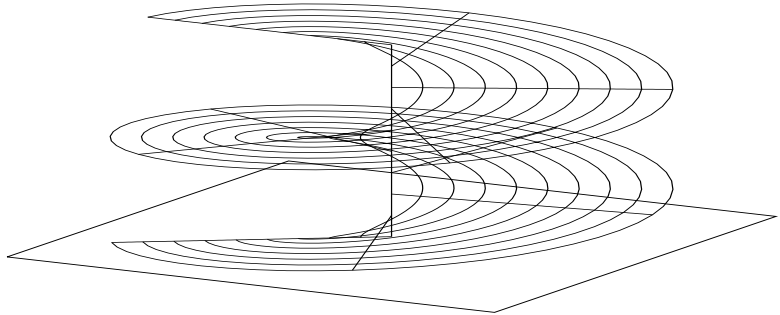}}
\end{center}
\caption{
Schematic view of two different manifolds which we consider as
embeddings for the chiral $O(2)$ model: the $R^2$ plane spanned by
cartesian components $(\sigma,\pi)$, and the angular 
$S^1\times R^1$ manifold spanned by the radial and angular coordinates
$(r,\phi)$
with its different sheets for the illustrative purpose 
pulled apart in vertical direction.}
\label{fig1} 
\end{figure}

To avoid the above conclusion it would be desirable to
combine the attractive features of the $O(4)$ model 
with the topological advantage of the nonlinear $\sigma$-model.
The decisive feature is the angular nature of the chiral field.
It is well known (see e.g. the discussion in~\cite{Polyakov}) 
that to specify an action in the continuum limit is not sufficient 
to fully specify a model. 
Additionally, the nature of the variables (continuous or
periodic) must be specified on physical grounds. In our case it is 
baryon number conservation which therefore requires embedding
the linear $\sigma$-model into the
$S^3 \times R^1$ manifold as natural extension of the
nonlinear $\sigma$-model, in contrast to the commonly used $R^4$ 
embedding. 
This means to insist on the angular 
nature of the chiral fields, i.e. on the factorization
\be
\label{Phi}
\Phi= R(\bx,t) U(\bx,t),~~~~~~~~~~~~~(U \in SU(2)). 
\ee
where the unitary part $U$ as usual is parametrized by
three chiral angles $\alpha(\bx,t)$,$\beta(\bx,t)$,$\gamma(\bx,t)$ 
\be
\label{abc}
U=\exp(i \;\gamma\;\btau \cdot \hat{\bphi}),
~~~~~~~\hat{\bphi}=(\cos\alpha\sin\beta,\sin\alpha\sin\beta,\cos\beta).
\ee
This generates a discrete set of different continuous manifolds of
angles $\alpha,\beta,\gamma$ which all describe the 
vacuum configuration $U=1$ which the field must approach 
at spatial infinity $|\bx|\to \infty$.
They constitute distinct sets of boundary conditions for field 
configurations with finite energy and definite winding number. 
(In the plain $R^4$ embedding it is not
possible to distinguish these different boundary conditions). 
As long as one of those boundary conditions is held fixed during the
time evolution, the winding number as defined in (\ref{B}) 
is conserved while the field $\Phi$ 
can explore the manifold $S^3 \times R^1$ without any further
constraint. Thus identification of $B$ with baryon number
is meaningful.

Whenever a field configuration $\Phi(\bx,t)$ ( which we shall briefly 
call a 'trajectory') at some point in space and time acquires the
value $\Phi=0$, i.e. if a trajectory moves across the origin $R=0$,
the angular fields will jump by multiples of $\pi$. With boundary
conditions on the angular fields fixed during the time evolution, this
cannot happen, therefore trajectories will not be able to 
move across the origin and curled-up configurations 
will not be able to unwind by crossing the origin.
It is, however, not necessary to enforce this by a constraint
on $R$, or by a singularity
of the potential at the origin. It is enforced by the boundary conditions
on the angular fields alone and reflects the fact that
in the $S^3 \times R^1$ manifold the origin $R=0$ as a highly singular
branching point is excluded.

One may ask how this will affect the time evolution of trajectories
which move according to the standard action of the $O(4)$ model. 
Ideally, without any damping mechanism, continuous trajectories moving
across the origin will develop hairpin slings which tie them to
the origin and serve to satisfy the angular boundary conditions, but do
not otherwise affect the motion of the trajectories. In a real cooling
process, however, all fluctuations are subject to dissipative damping
which will be especially efficient for the high Fourier components of
the slings. As a result, different embeddings will lead  
to different final configurations, irrespective of the
precise nature of the dissipative mechanism. 

Due to their topological equivalence the 1+1 dimensional $O(2)$ model
can serve as a transparent illustration
for the effects we may expect in the 3+1 dimensional $O(4)$ model.
In the following sections we discuss some typical aspects:
in sect.II we compare in angular and cartesian embedding 
the motion across the origin for very simple
trajectories without and with potential ; in sect.III we determine
the classical stable configurations and we show how the angular
boundary conditions stabilize nontrivial solutions which collapse
in the usual $R^2$ embedding to a point; finally, in sect.IV,
we follow the time evolution of random initial configurations 
after a quench.

\section{Motion of trajectories in different embeddings}

For the case of the 1+1 dimensional $O(2)$ model the two
different embeddings are easily visualized: in the usual $R^2$ embedding
the two-component chiral field $\Phi=(\sigma,\pi)$ lives on the 
$R^2$ plane spanned by cartesian components $-\infty<\sigma<\infty,
-\infty<\pi<\infty$. The origin is naturally included without any singularity.
The angular $R^1\times S^1$ embedding in radial and angular variables
($0<r<\infty, -\infty<\phi<\infty$) 
\be
\label{sigpi}
\sigma(x,t)=r(x,t)\;\cos\phi(x,t),~~~~~~~~\pi(x,t)=r(x,t)\;\sin\phi(x,t)
\ee
creates a manifold which consists of multiple sheets tied together at
the origin $r=0$ as branching point, like the Riemann sheets of the
complex plane. This is illustrated in fig.1 which schematically shows
the $R^2$ plane and the $R^1\times S^1$ manifold with the different
sheets pulled apart in vertical direction.
Imagine a path embedded in the $R^1\times S^1$ manifold
connecting two points which differ in $\phi$ by $2\pi$. Its projection
into the $R^2$ plane is a closed loop which can be contracted to a
point. But, evidently, with the endpoints fixed, the path on the
angular manifold cannot be pulled across the origin and thus cannot be
reduced to a point. 

Specifically, let the discrete manifold 
\be
\label{nvac}
r=f_\pi,~~~~~\phi=n 2 \pi,  
\ee
($n$ integer) represent an infinite set of distinct degenerate vacua.
Choosing for a trajectory $\Phi(x,t)$ fixed boundary conditions 
$\phi(\pm \infty,t)=2 \pi\: n_\pm$ fixes
the net winding number $B$ 
\be
\label{bnum}
B=\frac{1}{2\pi}\int_{-\infty}^{\infty} \frac{\partial}{\partial x}
\phi(x,t) dx = n_+-n_-,
\ee
so that identification of $B$ with conserved 'baryon number' is 
possible.

On the other hand, in the cartesian $R^2$ embedding 
all field configurations with boundary conditions 
\be
\label{bcsig}
\sigma(-\infty,t)=\sigma(+\infty,t)=f_\pi,~~~~~~~~~~~
\pi(-\infty,t)=\pi(+\infty,t)=0,
\ee
represent closed loops in the $\sigma$-$\pi$ plane
which can be contracted continuously into the unique vacuum
configuration $\sigma\equiv f_\pi, \pi\equiv 0$ and the integral
(\ref{bnum}) changes by one unit whenever the loop is pulled across
the origin of the $\sigma$-$\pi$ plane.
\begin{figure}[h]
\begin{center}
\leavevmode
\hbox{\epsfysize=5.8truecm \epsfbox{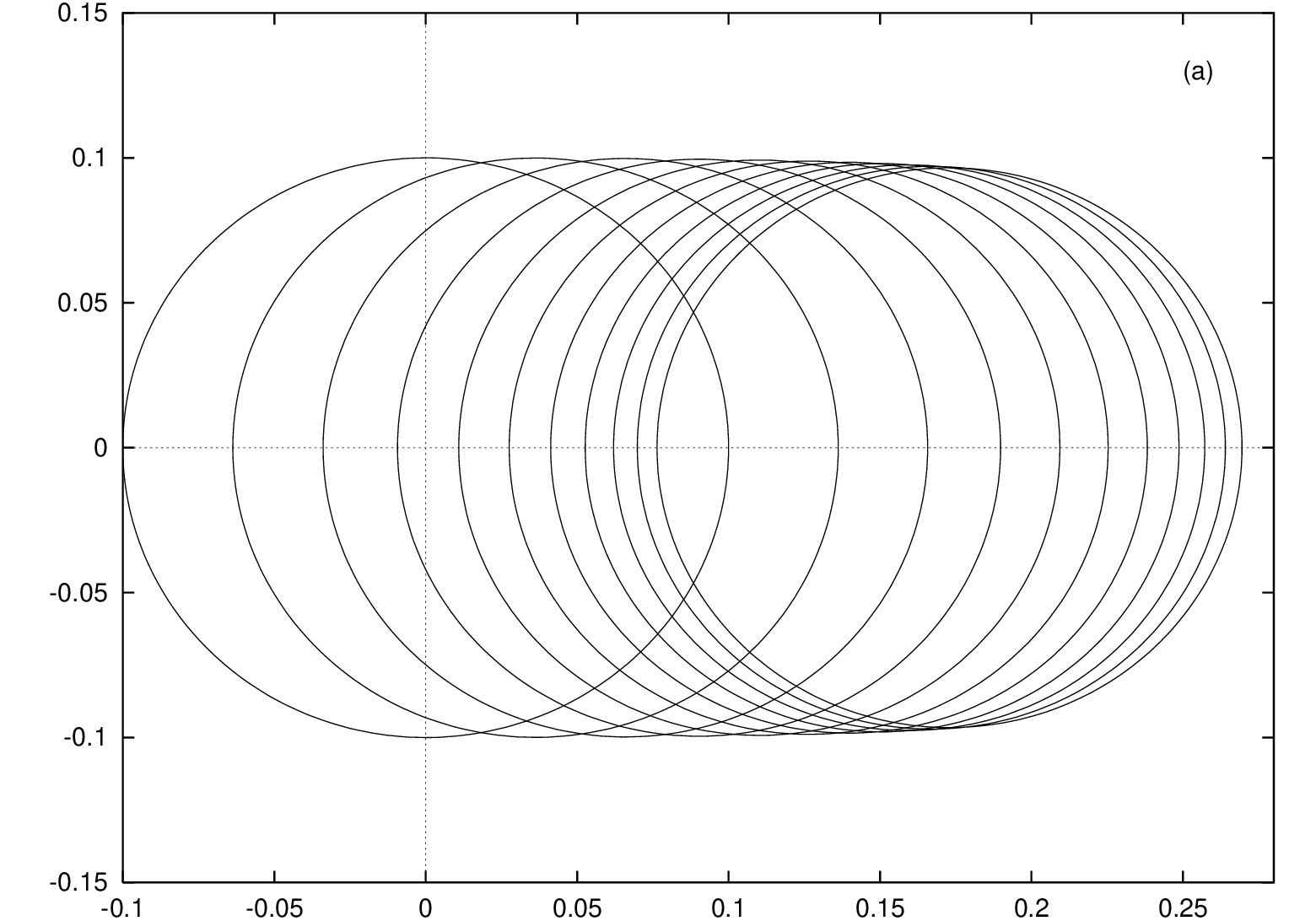} 
\epsfysize=5.8truecm \epsfbox{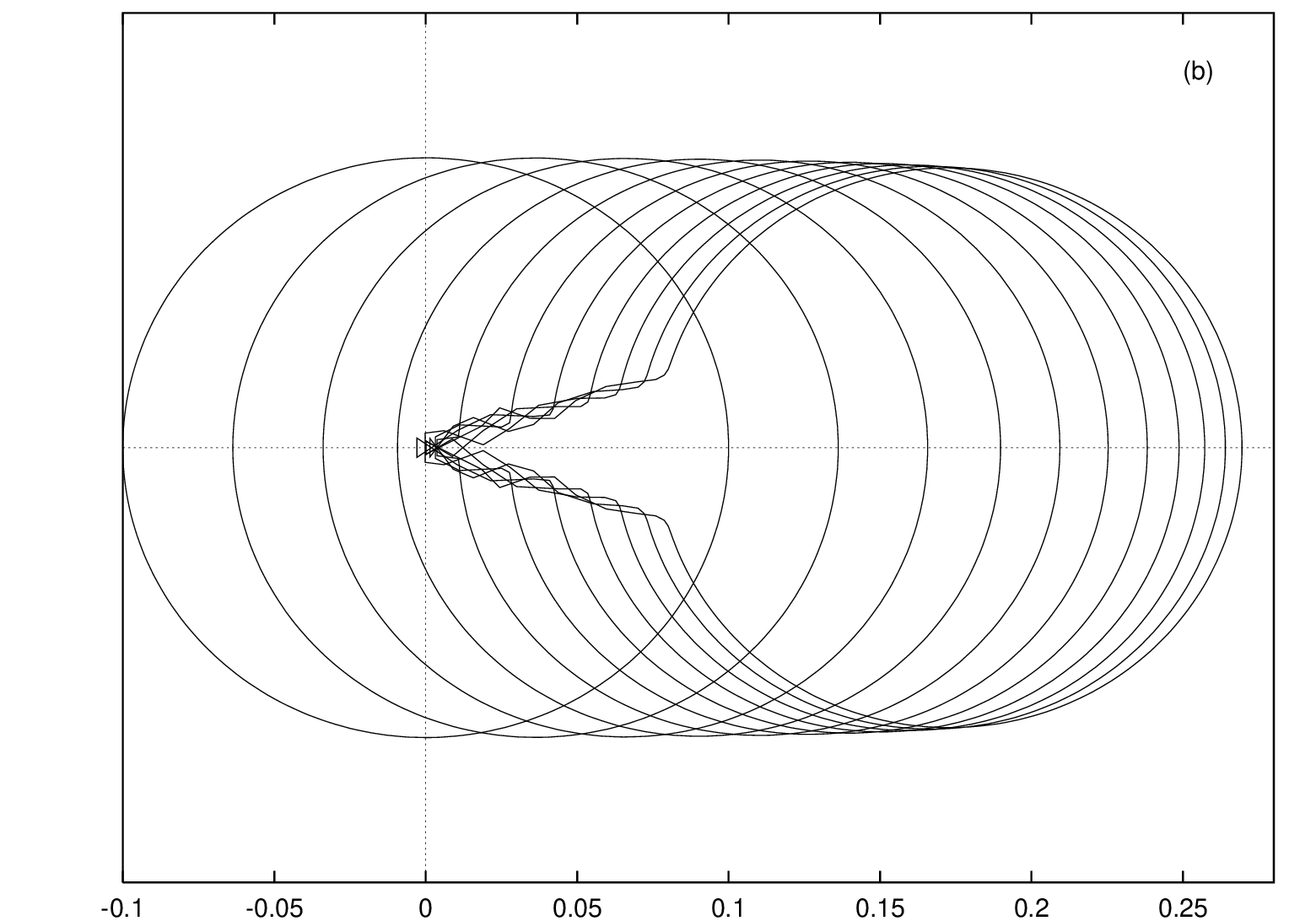}}
\end{center}
\caption{Free motion of a circular trajectory with radius $r=0.1$ 
across the origin (with small damping)
in 
(a) $R^2$ and in (b) $S^1 \times
R^1$ embedding of the $O(2)$ model.}
\label{fig2} 
\end{figure}

As a most simple example let us consider the time evolution of
a circular string moving to the right with a constant initial velocity
across the origin in the absence of any potential. So the underlying
Lagrangian density simply is 
\be
\label{L0}
{\cal L}_0(\Phi) = \frac{1}{2} \partial_\mu \Phi \cdot \partial^\mu \Phi.
\ee
In 'cartesian' form the equations of motion are\footnote{In the 
following $x$ and $t$ denote dimensionless
 spacetime variables. Where numbers are given, space and time are
 measured in units of the inverse physical pion decay constant of 93 MeV.}
\be
\label{xy}
\ddot{\sigma}=\sigma'' -\dot{\sigma},~~~~~~~~~~ \ddot{\pi}=\pi'' -\dot{\pi}
\ee
with boundary conditions
$   \pi(-\infty,t) =\pi(+\infty,t) =0 $,
and in 'angular' form 
\be
\label{rs}
\ddot{r}=r'' - r(\phi'^2-\dot{\phi}^2) - \dot{r},~~~~~~~~~
\ddot{\phi}=\phi''+\frac{2}{r}(r'\phi'-\dot{r}\dot{\phi}) -\dot{\phi}
\ee
with boundary conditions
$  \phi(-\infty,t)=0,~~ \phi(+\infty,t) =2\pi$.
In both sets of differential equations the same
dissipative terms (linear in the first time-derivatives)
have been added to smoothen fluctuations which build up in the 
'angular' parametrization. The effect of the dissipative terms 
can be seen in both cases
as slowing down the translational motion. Figures 2a,b show the 
same ten consecutive equal time steps in a numerical evolution 
in the $\sigma$-$\pi$ plane (a) and in the $r$-$\phi$ manifold (b).
Evidently, as consequence of the topological constraint in case (b) 
the circular trajectory after crossing the origin 
drags behind it a tail of fluctuations, their precise form depending on
mesh and damping details.

\begin{figure}[h]
\begin{center}
\leavevmode
\hbox{\epsfysize=5.8truecm \epsfbox{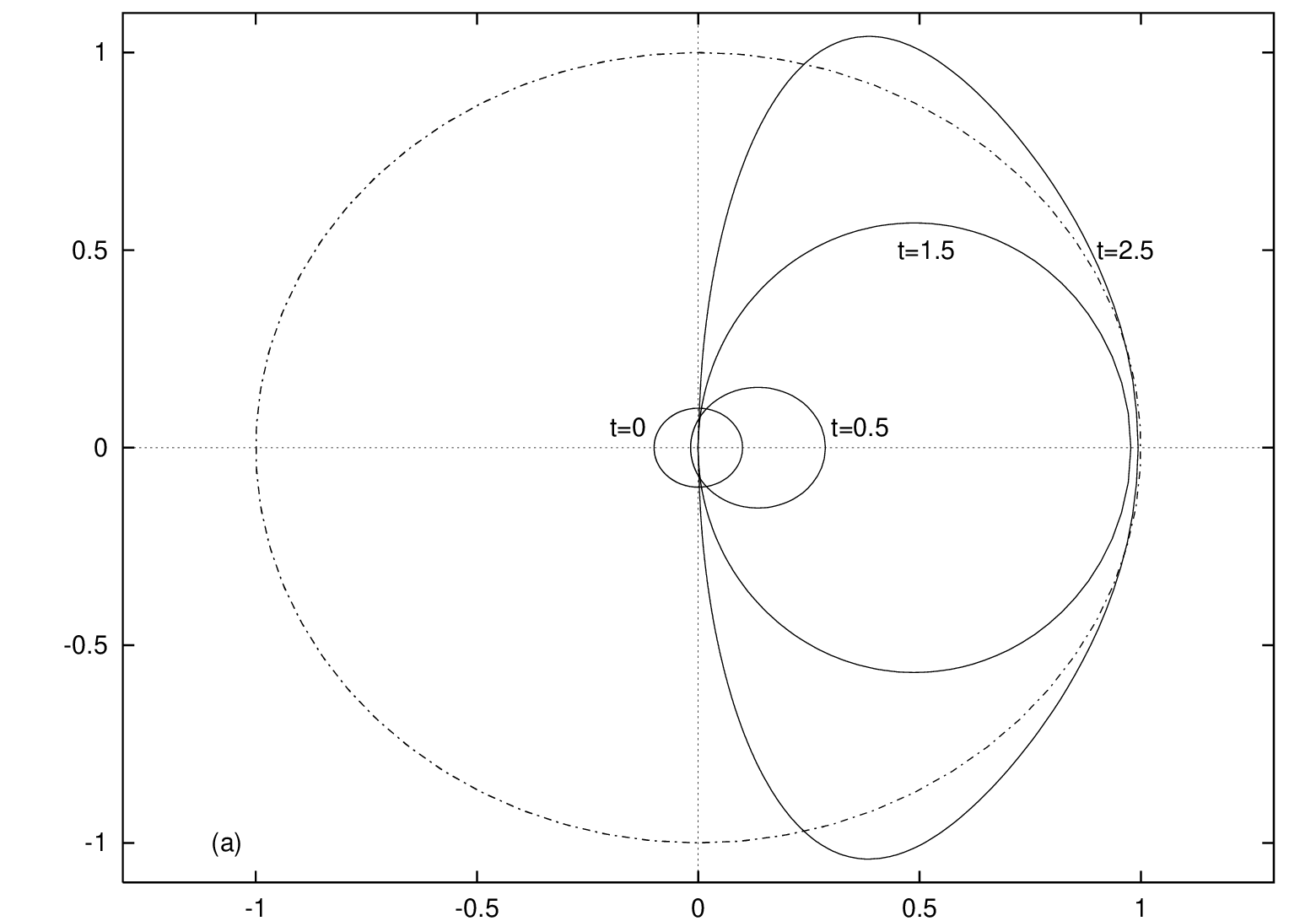} 
\epsfysize=5.8truecm \epsfbox{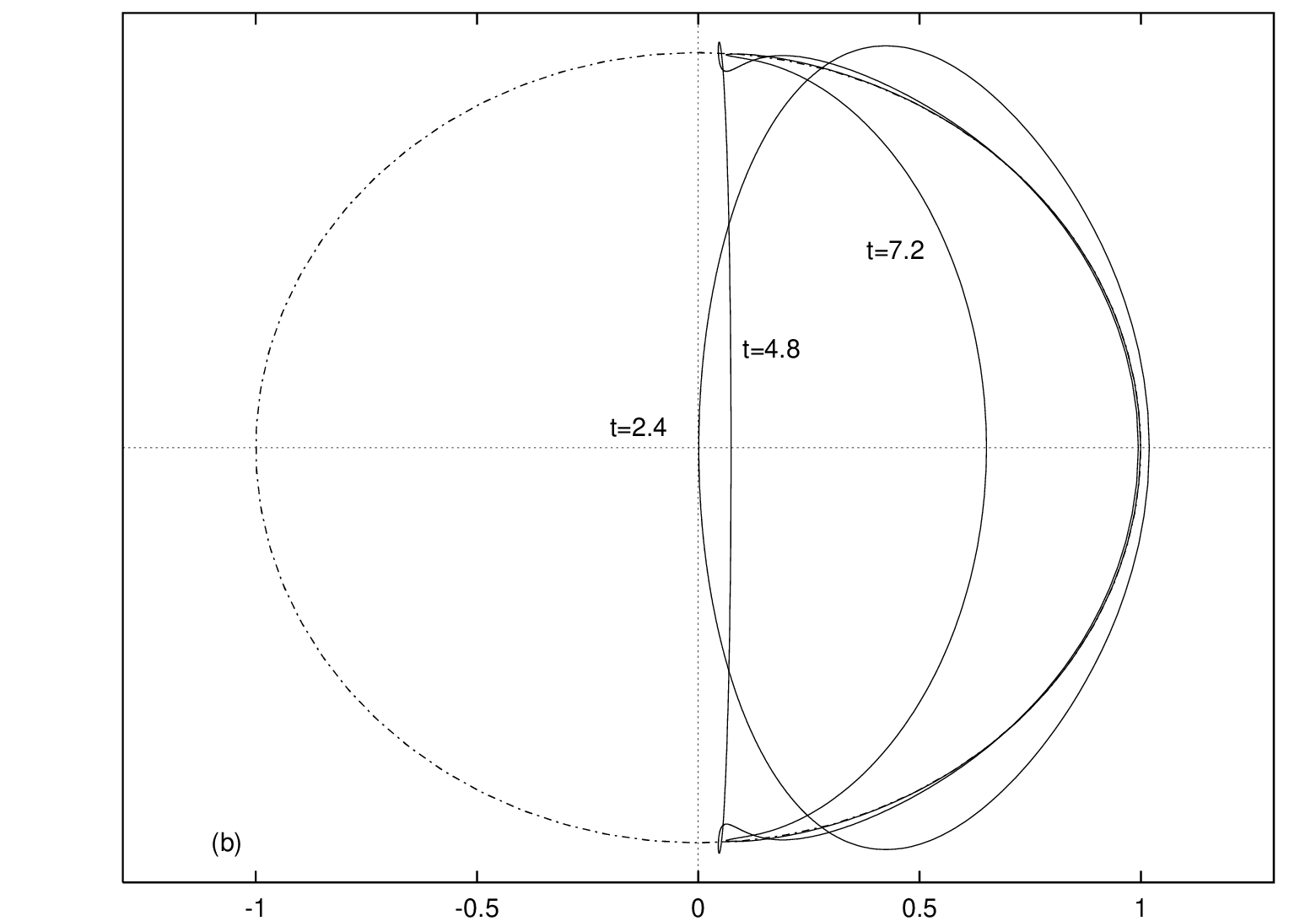}}
\end{center}
\begin{center}
\leavevmode
\hbox{\epsfysize=5.8truecm \epsfbox{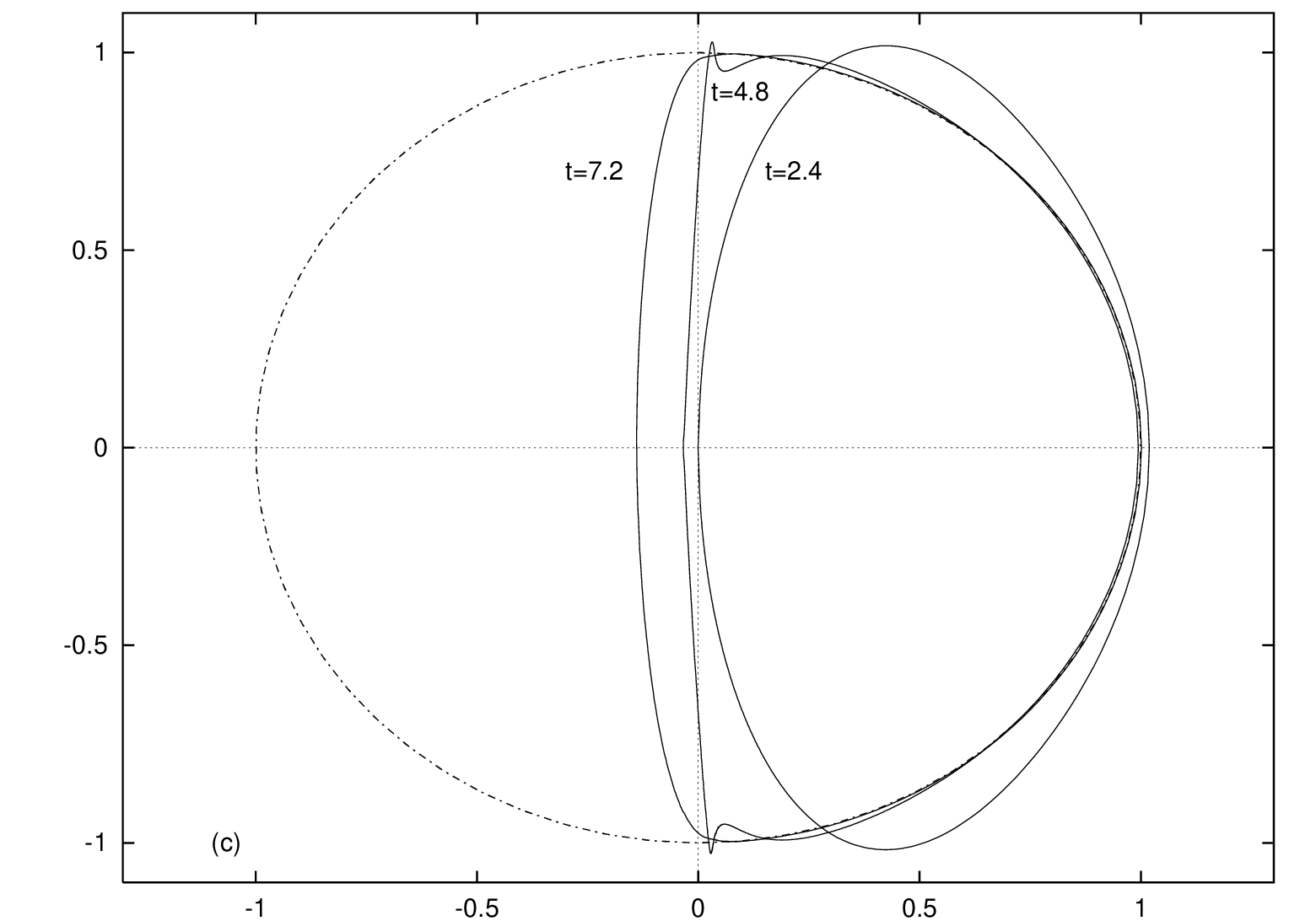} 
\epsfysize=5.8truecm \epsfbox{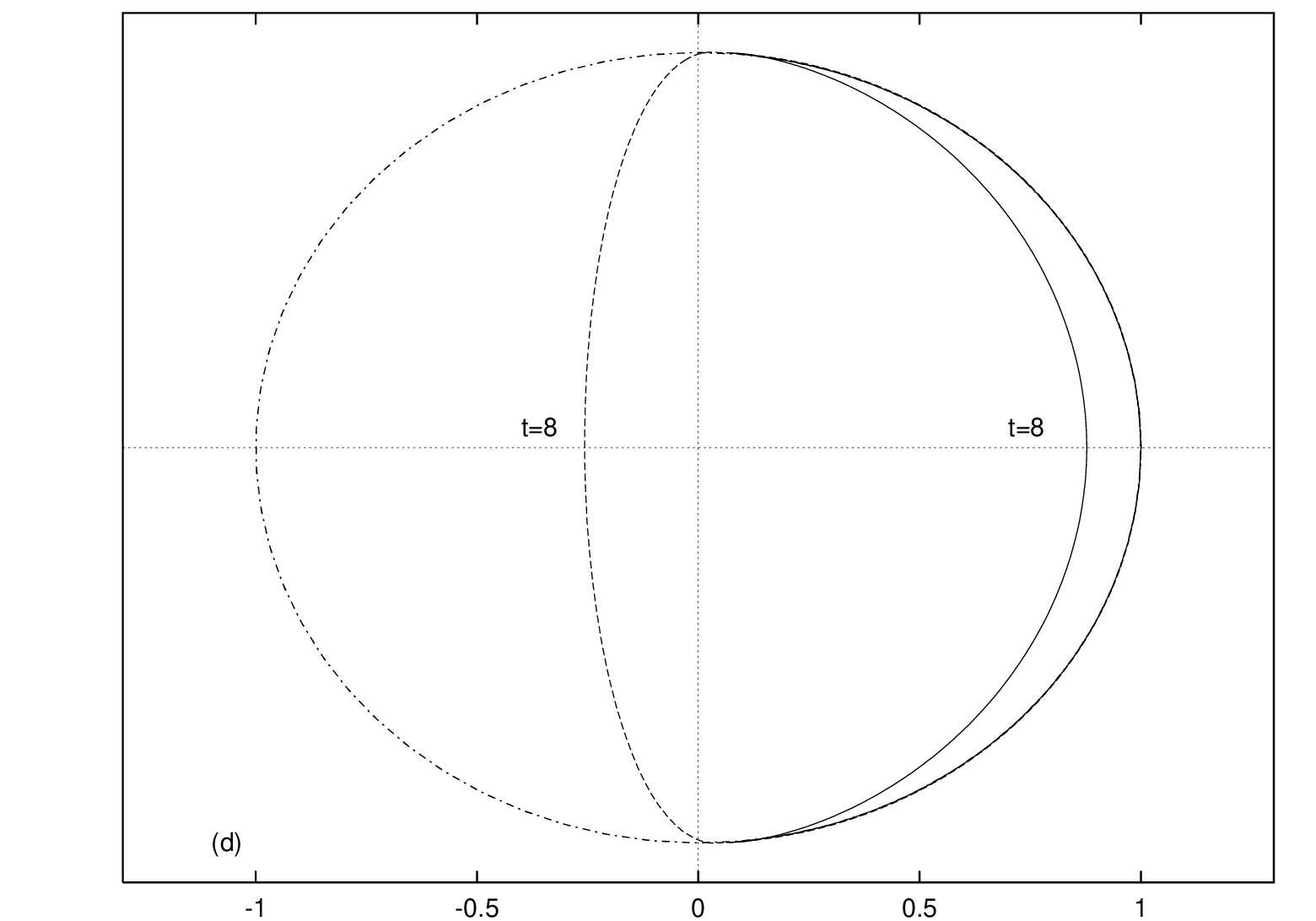}}
\end{center}
\caption{Evolution of an initially small circular trajectory 
with radius $r=0.1$ (moving with constant initial velocity as in fig.2)  
in the potential (\ref{pot}),
in $R^2$ and in $S^1 \times R^1$ embedding: 
Up to $t=2.5$ both evolutions coincide (a);
(b) shows the further cartesian evolution for $t=2.4,~4.8,~7.2$;
(c) shows the angular evolution at the same time steps; in (d) both
configurations are compared at $t=8$ (full line: cartesian, dashed line:
angular evolution). The dash-dotted circle denotes the bottom of 
the potential (in units of $f_\pi$).} 
\label{fig3} 
\end{figure}

Adding now to ${\cal L}_0(\Phi)$ the standard bell-shape potential 
\be
\label{pot} 
{\cal V} (\Phi) = \frac{\lambda}{4} (\Phi^2 - f_\pi^2)^2 
\ee
we may again in the two different manifolds follow 
the time evolution of the same circular
trajectory (initial radius $r=0.1 f_\pi$ with constant initial velocity
in positive $\sigma$ direction) on the way to its final
configuration located at the bottom of the well $r=f_\pi$. 
Up to the point where the trajectory meets
the origin both evolutions are identical. While the left part of the
circle is slowing down as it moves up towards the center top of the 
potential, the right half is rapidly accelerated downhill towards the
bottom of the well. (Fig.3a shows the timesteps $t$=0, 0.5, 1.5, 2.5).
In the $\sigma$-$\pi$ evolution the left part then freely moves across
the origin and follows the right half downhill into the same section of
the bottom circle. (Fig.3b shows the timesteps $t$=2.4, 4.8, 7.2).
In contrast to this, the $r$-$\phi$ evolution is held up at the origin
and then slowly accelerated downhill on the left side of the well
(fig.3c shows the same timesteps as fig.3b). Fig.3d shows both
trajectories at time $t=8$, where the right halfs both have settled
at the bottom $r=f_\pi$, the left part of the $\sigma$-$\pi$ trajectory
is approaching the bottom on the same side, while the left part of the
$r$-$\phi$ trajectory is still moving downhill to complete after a few
further timesteps the full circle which conserves the winding number of
the initial $t$=0 configuration.   
Evidently, we can have with identical dynamics a dramatic difference 
of the evolutions in the different manifolds. 
Of course this only happens if the initial
velocity is sufficient for the trajectory to overcome the center top 
of the potential. For smaller velocities both trajectories will 
evolve identically and both conserve winding number.

\section{Stable solitons}

The defects which carry winding number are not necessarily stable
static configurations. Therefore, if their winding number is not 
topologically protected unstable defects will unwind and disappear. 
On the other 
hand, unstable topological defects will degenerate into spatially isolated
singularities. It is therefore natural to expect that the stabilization
mechanism of solitonic configurations will play an essential role for
the evolution of trajectories.

In the 3+1-dimensional $O(4)$-model soliton stabilization requires terms
of higher chiral order, like the fourth-order Skyrme term~\cite{Skyrme}.
The spatial extent of the angular chiral soliton profile scales
as $(f_\pi e)^{-1}$ where the Skyrme parameter $e$ remains essentially
unchanged with increasing temperature. 
The angular winding length $L_w\sim f_\pi^{-1}$ 
as the characteristic length scale for stable field configurations
then increases with increasing temperature. The
radial bag with increasing $L_w$ gets very shallow, 
i.e. $R(\bx)$ stays very close to $f_\pi$ as $f_\pi$ 
approaches zero with increasing temperature. 
So the static soliton dissolves
together with restoration of chiral
symmetry. But the baryon number defined through the
choice of boundary condition in the $S^3 \times R^1$ embedding 
continues to be carried by the hot chiral gas even if the soliton has 
dissolved into chiral fluctuations around the vanishing mean value of
the condensate.

It is nice to observe that also in the $O(2)$ model there is a simple
way to destabilize static solitons and thus to choose any desired
angular winding length. But unlike in the $O(4)$ case, here the mechanism
is tied to explicit breaking of the $O(2)$ symmetry. Evidently, the static
configuration 
\be
\label{static}
r=f_\pi,~~~ \phi=B\pi(1+x/L),
\ee
($2L$ is the size of the spatial box) which rests at
the bottom of the well and winds $B$ times 
around the center, is stable
for unbroken symmetry. It is also evident, that if we add a
symmetry-breaking term $H\sigma$ to the potential
\be
\label{brokpot} 
{\cal V} (\Phi) = \frac{\lambda}{4} (\Phi^2 - f^2)^2 - H \sigma \; .
\ee
there will be a maximal value of $H$ which characterizes the onset
of instability of static configurations with nonvanishing winding number.
In order to keep the minimum of ${\cal V}$ for finite $H$ at 
$\Phi^2 = f^2_\pi$ we define
\be
f^2 = f^2_\pi - \frac{H}{\lambda f_\pi} \;.
\ee
The stable solitons correspond to bound trajectories of a classical
point particle moving in the potential $-{\cal V}$, starting from
the maximum of $-{\cal V}$ and returning to it after an infinite time.
These configurations sling 
around the hat of the potential 
${\cal V}$ close to (but inside of) the bottom of
the valley. Balancing centrifugal forces and potential gradient for
such a classical particle provides the limiting value of $H$: 
stable solitons exist, as long as the
inequality
\be
\label{ineq}
\frac{H}{\lambda f_\pi^3} < 0.047
\ee
is satisfied.  In terms of $\sigma$- and $\pi$-masses
\be
\label{masses}
m_\pi^2=\frac{H}{f_\pi},~~~~~~~m_\sigma^2=2 \lambda f_\pi^2 + m_\pi^2
\ee
we find from (\ref{ineq}) approximately $m_\sigma > 6.5\; m_\pi$, as
stability condition.

\begin{figure}[h]
\begin{center}
\leavevmode
\hbox{\epsfysize=5.8truecm \epsfbox{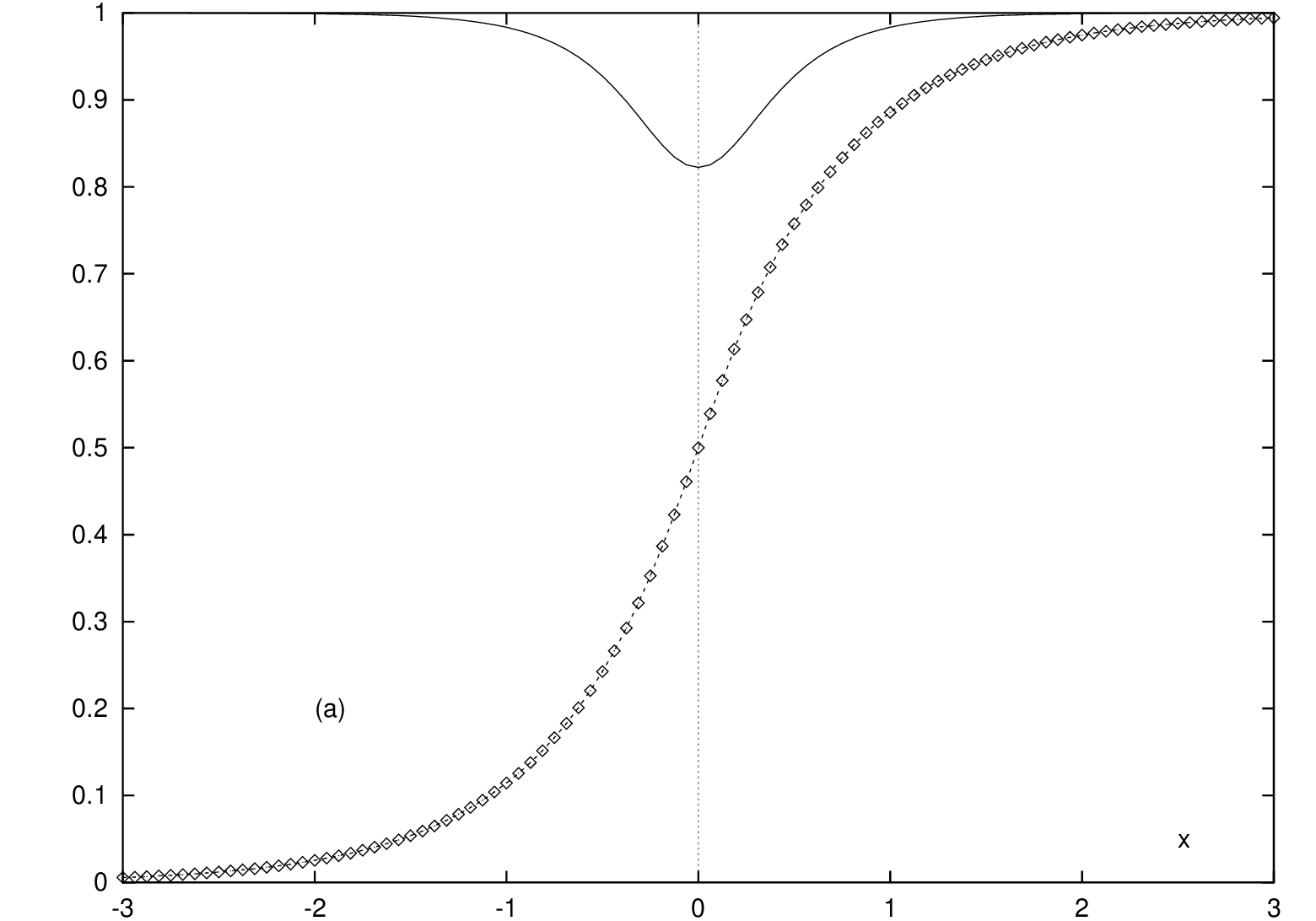} 
\epsfysize=5.8truecm \epsfbox{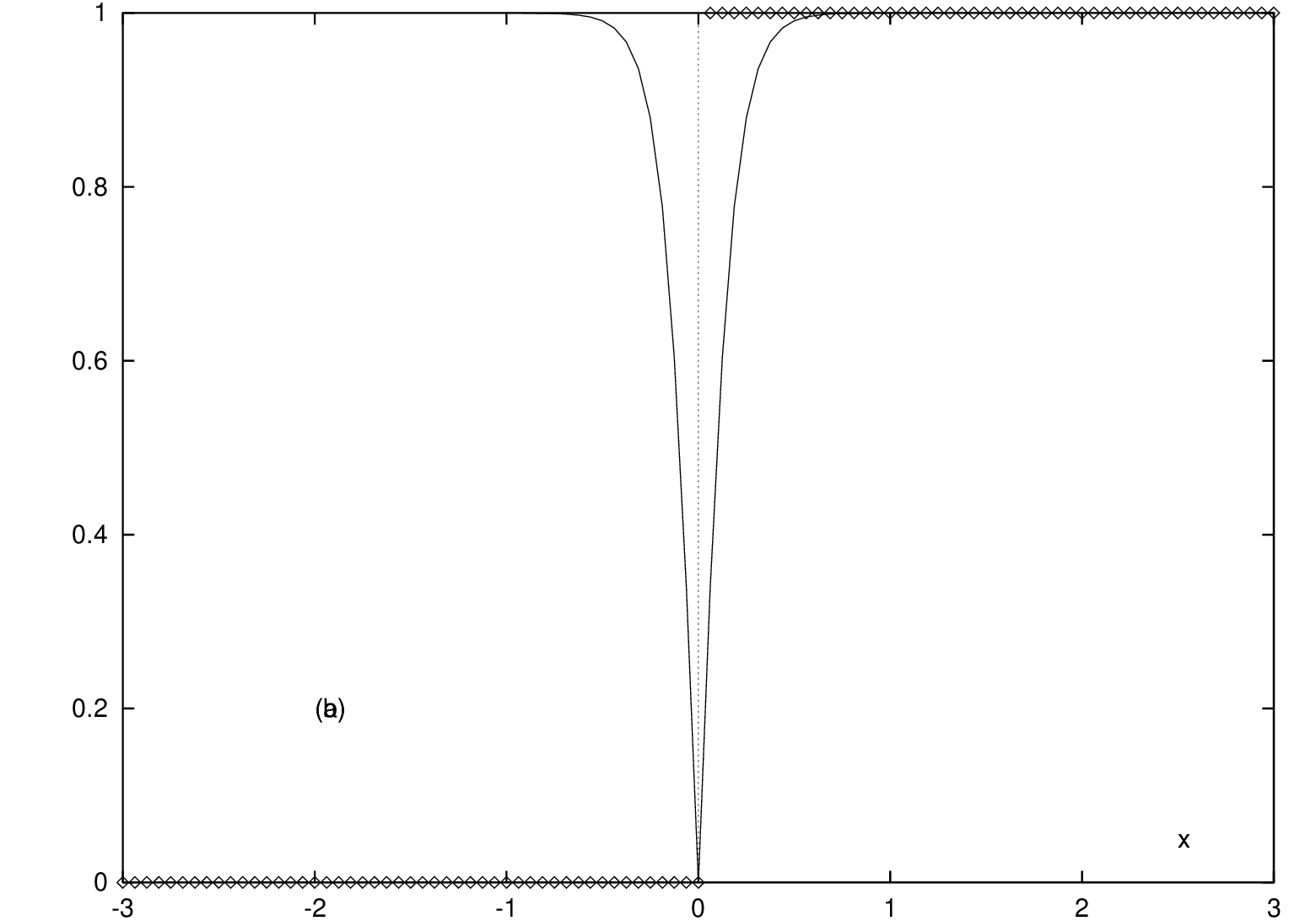}}
\end{center}
\caption{Profiles of stable solitons in the $O(2)$ model ($m_\sigma=1000$ MeV,
$f_\pi=93$  ) for two different values of $m_\pi$. In (a) $m_\pi=140$
MeV is below, in (b) $m_\pi=200$ MeV is above the critical value of
$m_\sigma/6.5=154$ MeV. The full line is the radial bag profile
$r(x)$ (in units $f_\pi$), the linespoints show the corresponding
angular profiles $\phi(x)$ (in units of $2\pi$). For $m_\pi>154$ MeV
$\phi$ is a step function. }
\label{fig4} 
\end{figure}

The angular winding length $L_w\sim (\phi'(x=0))^{-1}$ 
(i.e. the inverse of the gradient of the angular field near the center
of the soliton) defines the typical spatial extent of the stable
soliton. In the symmetry limit $H\to 0$ $\phi(x)$ approaches
$\pi(1+x/L)$, so in this limit the soliton completely occupies 
the spatial box of length $2L$. With increasing $L_w$ the radial bag
$r(x)$ gets very shallow, i.e. $r(x)$ stays very close to $f_\pi$.

Suppose that $H$, $\lambda$ and $f_\pi$ satisfy condition (\ref{ineq})
for $T=0$  and suppose we have a 'baryon' represented by a stable 
trajectory with $B=1$. With increasing temperature $T$,
the inclusion of loop corrections into a renormalized ${\cal L}$
will lead to a decrease in $f_\pi^2(T)$ while
the coupling constant $\lambda$ and $H$ remain unrenormalized. 
This causes the soliton radius to decrease. At a
certain temperature the value of $H/(\lambda f_\pi^3)$ will exceed 
the critical stabilizing value (\ref{ineq}), and in an $R^2$ embedding
the sling will collapse to the vacuum point $\sigma\equiv f_\pi,\pi\equiv 0$, 
and the 'baryon' has disappeared.

If we repeat these considerations in the $S^1 \times R^1$ embedding
starting with a stable trajectory at $T=0$ with boundary conditions
$\phi(-\infty,t)=0,\phi(\infty,t)=2\pi$, increase the temperature, 
then at the critical 
value (\ref{ineq}) for $f_\pi(T)$ the sling will now
collapse to a narrow hairpin around the origin
which connects the vacua (\ref{nvac}) with $n_-=0$ and $n_+=1$,
$\phi(x)=2 \pi \Theta(x)$. The radial field, however, still obeys a
nonlinear equation
\be
\label{radeq}
r''-\lambda r (r^2-f^2)+H=0
\ee
with nontrivial solutions satisfying
\be
\label{hairpin}
r(|x|\to 0) \to +0,~~~~~~~r(x\to\pm\infty) \to f_\pi-Ae^{-m_\sigma|x|}.
\ee 
which represent the 'bag' profile
of the remaining baryon with $B=1$. Although its angular winding length
$L_w$ has shrunk to zero, the radial bag still has a nonzero radius 
in coordinate space which
increases with temperature as long as the $\sigma$-mass decreases
(in this particular stabilization mechanism which is peculiar to
1 space dimension). In fig.4 soliton profiles are shown for two
values of $m_\pi$ above and below the critical value (\ref{ineq}). 
The stable soliton solutions above critical symmetry breaking
do not exist in the $R^2$ embedding of this model. 
    
\section{Evolution after a quench}

It is an interesting question to ask whether and 
how the topological aspects discussed above may affect the time
evolution of chiral fields after a quench where an initially hot
hadronic gas is rapidly cooled down across the chiral phase transition.
The initial configuration
therefore is characterized by an ensemble of randomly curled up
trajectories  which
reflect the unbroken symmetry of the hot chiral gas phase.  
The time evolution of each trajectory after the quench then is 
governed by the equations of motion in the low temperature
effective potential.
In the commonly used $R^4$ embedding the initial configurations 
therefore can be created as random Gaussian ensembles
centered around the origin $\Phi$=0~\cite{rw}. 
In the angular representation 
(\ref{Phi}),(\ref{abc}) of the
chiral fields this would correspond to uniform deviates in the 
three chiral angles 
$$
0\le\alpha\le 2\pi,~~~~0\le\beta\le \pi,~~~~0\le\gamma\le \pi.
$$ 
However, in $S^3\times R^1$ embedding 
the chiral angles extend over all values
$-\infty<\alpha,\beta,\gamma<+\infty$ and
chiral symmetry requires equal probability for all values.
So with each point of each trajectory all
integer multiples of $2\pi$ should also be included.
This implies that initial field configurations could curl arbitrarily
often back and forth around the origin from one lattice point to the
next, with only the net winding number $B$ of the configuration inside
the whole spatial box of radius $L$ fixed.
With our interpretation of the Jacobian
\be
b(\bx)=\frac{1}{2\pi^2}\sin\beta\sin^2\gamma \:
\frac{\partial (\alpha,\beta,\gamma)}{\partial (x,y,z)}
\ee 
as local baryon density such configurations apparently 
should be very much suppressed.

Instead, for high values of the temperature $T$ the initial
ensemble should be such that the correlation function at $t=0$
is characterized by a small correlation length $\xi_0\sim T^{-1}$ 
\be
\label{corrU}
\left\langle {\rm tr} ~U^\dagger(\bx+\br,0)U(\bx,0) \right\rangle
 \sim e^{-|r|/\xi_0}.
\ee
with dominant individual configurations characterized by
a small local baryon density. 
With $\xi_0 \ll L$ such configurations again wind many
times back and forth around the origin within the whole spatial box, 
but most of the phase differences
between neighbouring lattice points should be less than $2\pi$,

Starting from such initial configurations, 
together with a Gaussian deviate in $R(\bx,0) > 0$ around $R=0$,
in $S^3\times R^1$ embedding the growth of collective 
motion in radial direction towards 
the condensate cannot proceed (even for $B=0$) 
through unwinding random multiple twists around the origin
by moving the curled-up trajectories across the origin.
Furthermore, while the evolution of the radial part $R(\bx)$ 
towards the condensate is
driven by the slope of the potential $V(R)$, there is no
corresponding driving force to unwind the large amplitude
fluctuations in the angles $\alpha,\beta,\gamma$.  
However, inclusion of dissipative terms will relax also the 
random angular fluctuations into a 
quasistable ensemble of $B_+$ solitons and $B_-$ antisolitons 
(with $B=B_+-B_-$) distributed 
randomly within the spatial box. Although the
configuration then is characterized by a phase correlation length of
order $L_w$ it generally will contain many sections winding back and
forth around the origin.
Whether optimal
unwinding ($B_+=B$) can be achieved depends on the ratio of three
different length scales: the initial phase correlation length
$\xi_0$ ($\sim$ the lattice constant), the winding length $L_w$ of
stable (anti)solitons i.e. on the stabilization mechanism,
and on the size $L$ of the box. 
A necessary condition for optimal unwinding is $\xi_0 \ll L_w$, 
because otherwise the 
initial configuration will be close to some quasistable
soliton-antisoliton trajectory with comparable large mean square
deviations $\sim L/\xi_0$ in the chiral angles. 
If $\xi_0 \ll L_w \ll L$, the quasistable trajectory can be very
different from the initial incoherent fluctuations but still may
contain many multiply curled-up sections. 
Only if $L_w$ is comparable or
larger than $L/(B+2)$ then no stable soliton-antisoliton configurations 
fit into the box and the dissipative term can completely 
unwind all fluctuations. Of course, in all cases the net winding 
number $B$ will be conserved. 
Because the different length scales change with temperature,
the detailed study of the cooling process of the hot pion gas
in this extended $S^3 \times R^1$ frame (as compared to~\cite{EKK})
might provide interesting features of baryon-antibaryon formation. 

Again it is possible and illustrative to simulate such time evolutions 
in the $S^1 \times R^1$ $O(2)$ model.
As discussed above the initial configurations should reflect
the high value of the temperature $T$ through a small 
correlation length $\xi_0\sim T^{-1}$ characterizing the angular
correlation function at $t=0$
\be
\label{corr}
\left\langle e^{i\phi(x_0+x,0)}e^{-i\phi(x_0,0)}\right\rangle
=e^{-\frac{1}{2}\langle[\phi(x_0+x,0)-\phi(x_0,0)]^2\rangle}=e^{-|x|/\xi_0}.
\ee
This shows that the mean square net winding increases linearly with
$x$. On a lattice with lattice constant $a$ this corresponds 
to a random walk with $|x|/a$ steps with a mean
square step size of $2 a/\xi_0$.
Identifying $\xi_0$ with the lattice constant we obtain suitable
initial configurations for the angles $\phi(x,0)$ by random walks with
Gaussian deviate for mean square step sizes of order two, and fixed
value of total net winding $n$.
If $a\ll2L$ these configurations wind many
times back and forth around the origin, but most of the phase differences
between neighbouring lattice points are less than $2\pi$,
i.e. the local baryon density $\sim \phi'(x,0)/(2\pi)$
is small. The actual value of the total winding number $B=n_+-n_-$ 
of the configuration apparently 
is of minor importance for the general features of the evolution: 
as long as the field curls
many times back and forth around the origin, it is not 
essential at which multiple of $2\pi$ it ends.
\begin{figure}[h]
\begin{center}
\leavevmode
\vbox{\epsfxsize=13truecm
\epsfysize=6truecm \epsfbox{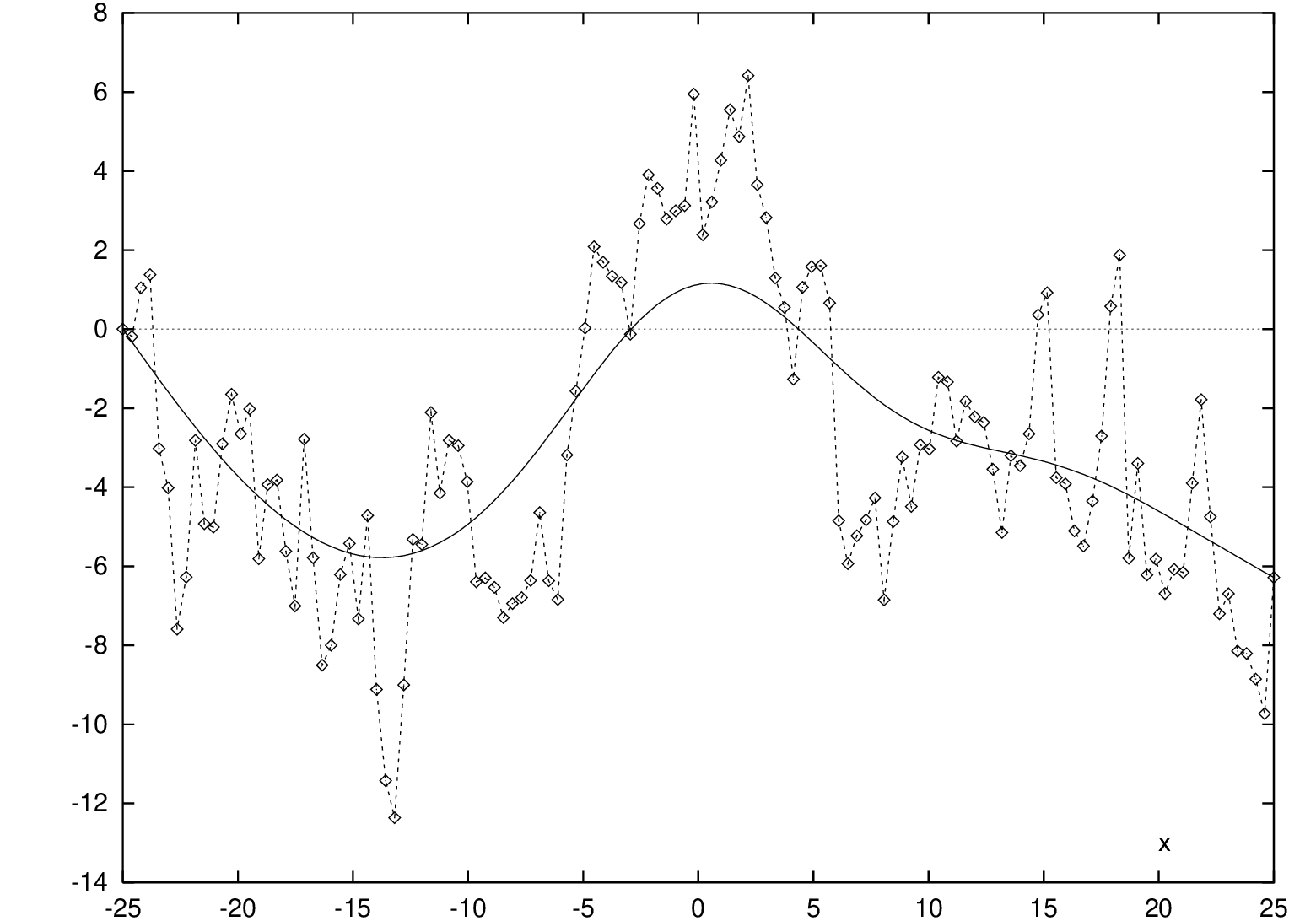}}
\end{center}
\caption{
The angular field $\phi$ at time $t=20$ (full line) after evolving
in the $S^1 \times R^1$ embedding of the $O(2)$
model from a random initial configuration $\phi(x,t=0)$ (linespoints)
for $m_\sigma=1000$ MeV, $f_\pi=93$, and no explicit symmetry
breaking ($ m_\pi=0$).}
\label{fig5} 
\end{figure}

As an illustration we consider the time evolution of an initial
configuration on 128 lattice points in a box $-L\le x\le L$ ($L=25$)
which consists of a Gaussian deviate in the radial field $r(x)$
around the origin $r=0$ and the angular configuration $\phi(x)$
given by a Gaussian random walk which begins at $\phi(-L)=0$ and
ends near $\phi(L)\approx -2\pi$ (cf. the linespoints in fig.5). 
In order to have integer $B$
we put $\phi(L)=-2\pi$ and keep both boundary values $\phi(-L)$,
$\phi(L)$ fixed for all times $t$. For a comparison with an
evolution in $R^2$ embedding we obtain the corresponding cartesian
initial configuration from (\ref{sigpi}) which then, of course, satisfies
$\pi(-L)=\pi(L)=0$. For the sake of our comparison we keep this  
boundary condition also fixed during the cartesian evolution.

The time evolution for $t > 0$ is governed by the potential
(\ref{pot}). (We take $m_\sigma=1000$ MeV and $f_\pi=93$  ).
Again we add dissipative terms in the equations of motion as in
(\ref{xy}),(\ref{rs}). So, in angular form we have 
\begin{eqnarray}
\label{eom}
\ddot{r}&=&r'' - r(\phi'^2-\dot{\phi}^2) 
-\lambda r (r^2-f^2)+H \cos \phi - \dot{r},   \nonumber \\
\ddot{\phi}&=&\phi''+\frac{2}{r}(r'\phi'-\dot{r}\dot{\phi}) 
-\frac{H}{r} \sin \phi -\dot{\phi}
\end{eqnarray}

Let us first consider the case of unbroken
symmetry, i.e. $m_\pi=0$. As we discussed in the previous section,
in this case the slings (\ref{static}) which rest at the bottom
of the well and wind $B$ times around the origin are stable classical
solutions which finally will be approached by any evolution in time.
The early phase ($t < 2$) of each evolution, 
angular or cartesian, is
characterized by a rapid increase in the average radius of the trajectory 
towards the bottom of the well. During this early phase the cartesian
trajectory crosses the origin repeatedly at different points 
in space and time  
with corresponding changes in net winding number $B$. In the following
($t < 10$) the evolutions are characterized by smoothing of the rapid 
random fluctuations. Finally, the
remaining long-range variations of the trajectories are slowly reduced
until finally the stable slings (\ref{static}) are approached ($t >
100$). 

\begin{figure}[h]
\begin{center}
\leavevmode
\hbox{\epsfysize=6.2truecm \epsfbox{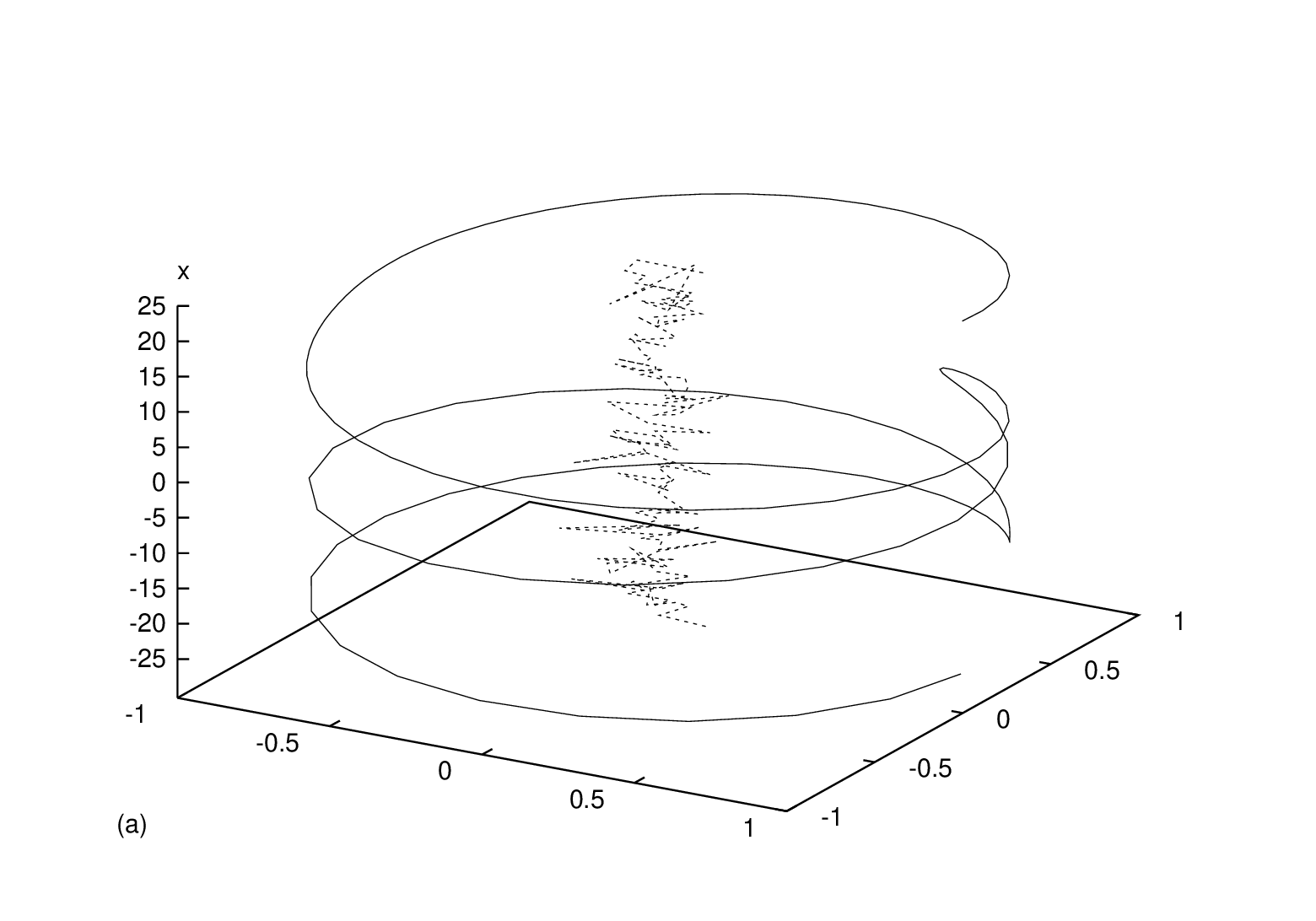}
\epsfysize=6.2truecm \epsfbox{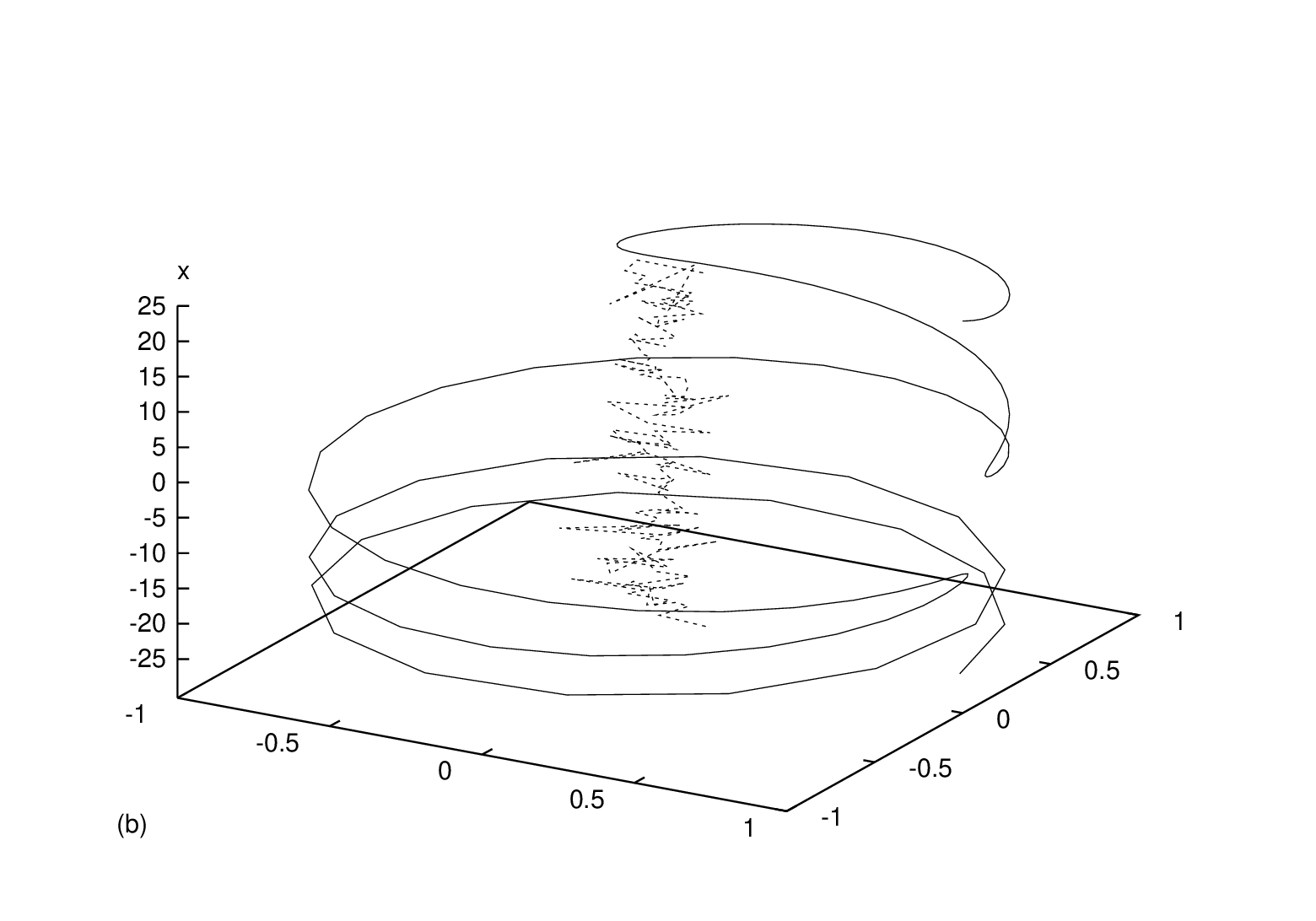}}
\end{center}
\caption{
Comparison between
trajectories in the $O(2)$ model at time $t=20$ after evolving 
in (a) $S^1 \times R^1$
and in (b) $R^2$ embedding from the same random initial configuration 
as in fig.5 (points connected by dashed lines near the center of the
$\sigma$-$\pi$ plane).
The plots show the fields $\sigma(x)$
vs. $\pi(x)$ vs. the (vertical) coordinate x. 
The angular field $\phi(x)$ in (a) coincides with the full line in fig.5. 
In both cases 
the trajectories begin and end at the vacuum values
$\sigma=1$, $\pi=0$, and
the radial field $r(x)$ has settled near the bottom of
the potential (\ref{pot}) at $r=1$ (in units of $f_\pi$). 
($m_\sigma=1000$ MeV, $m_\pi=0, f_\pi=93$).}

\label{fig6} 
\end{figure}

\begin{figure}[h]
\begin{center}
\leavevmode
\vbox{\epsfxsize=13truecm
\epsfysize=6truecm \epsfbox{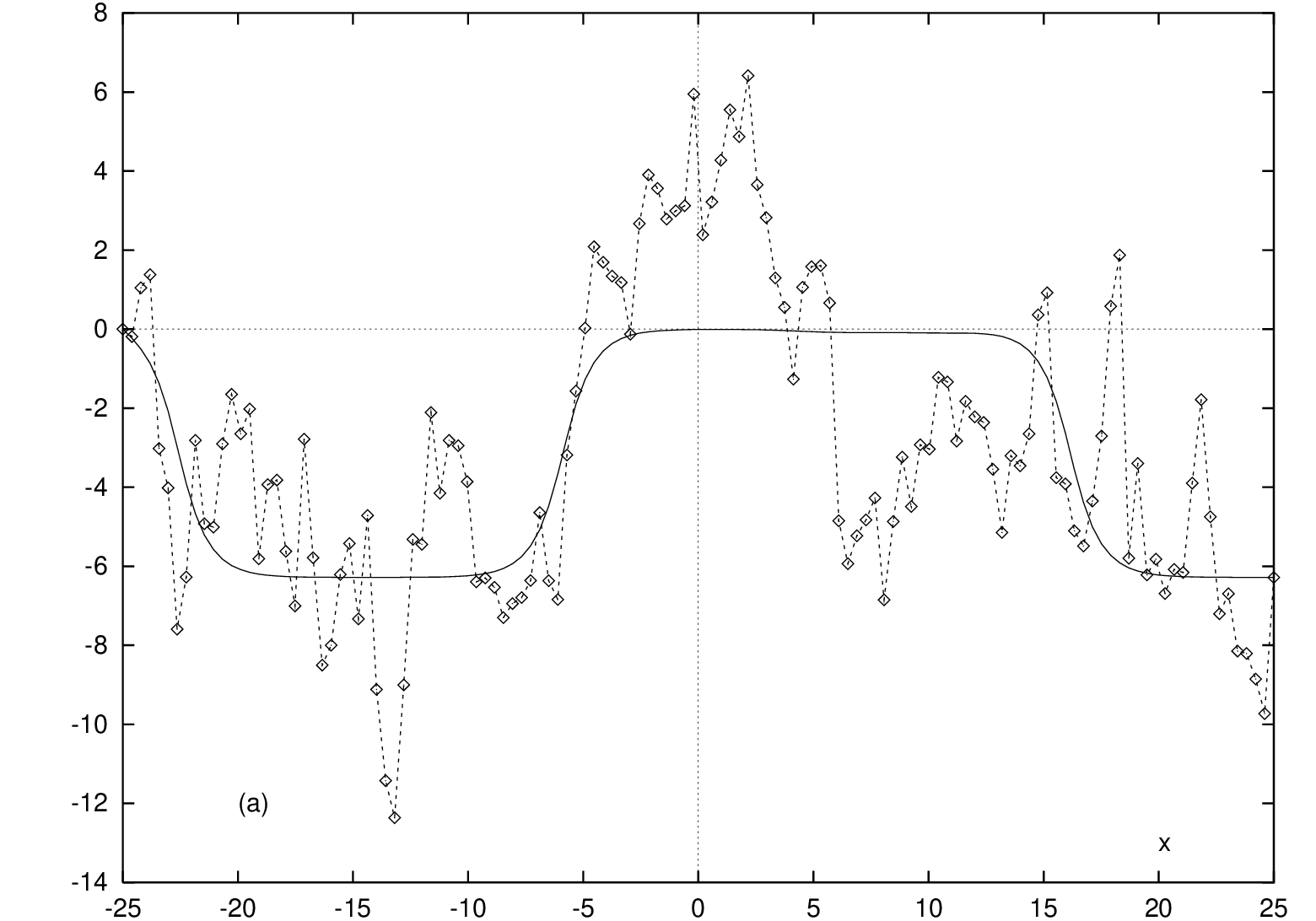}}
\end{center}
\begin{center}
\leavevmode
\vbox{\epsfxsize=13truecm
\epsfysize=6truecm \epsfbox{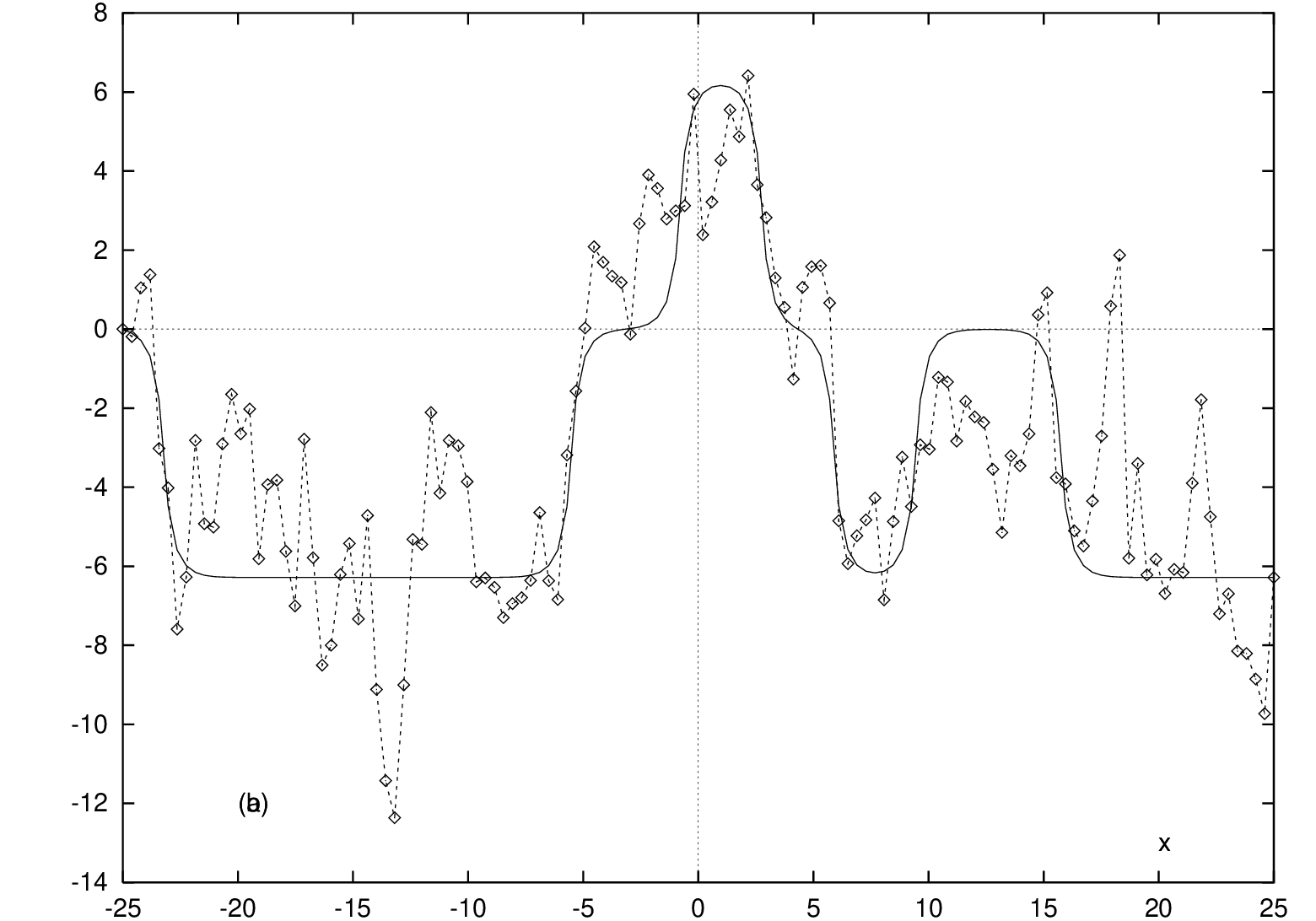}}
\end{center}
\begin{center}
\leavevmode
\vbox{\epsfxsize=13truecm
\epsfysize=6truecm \epsfbox{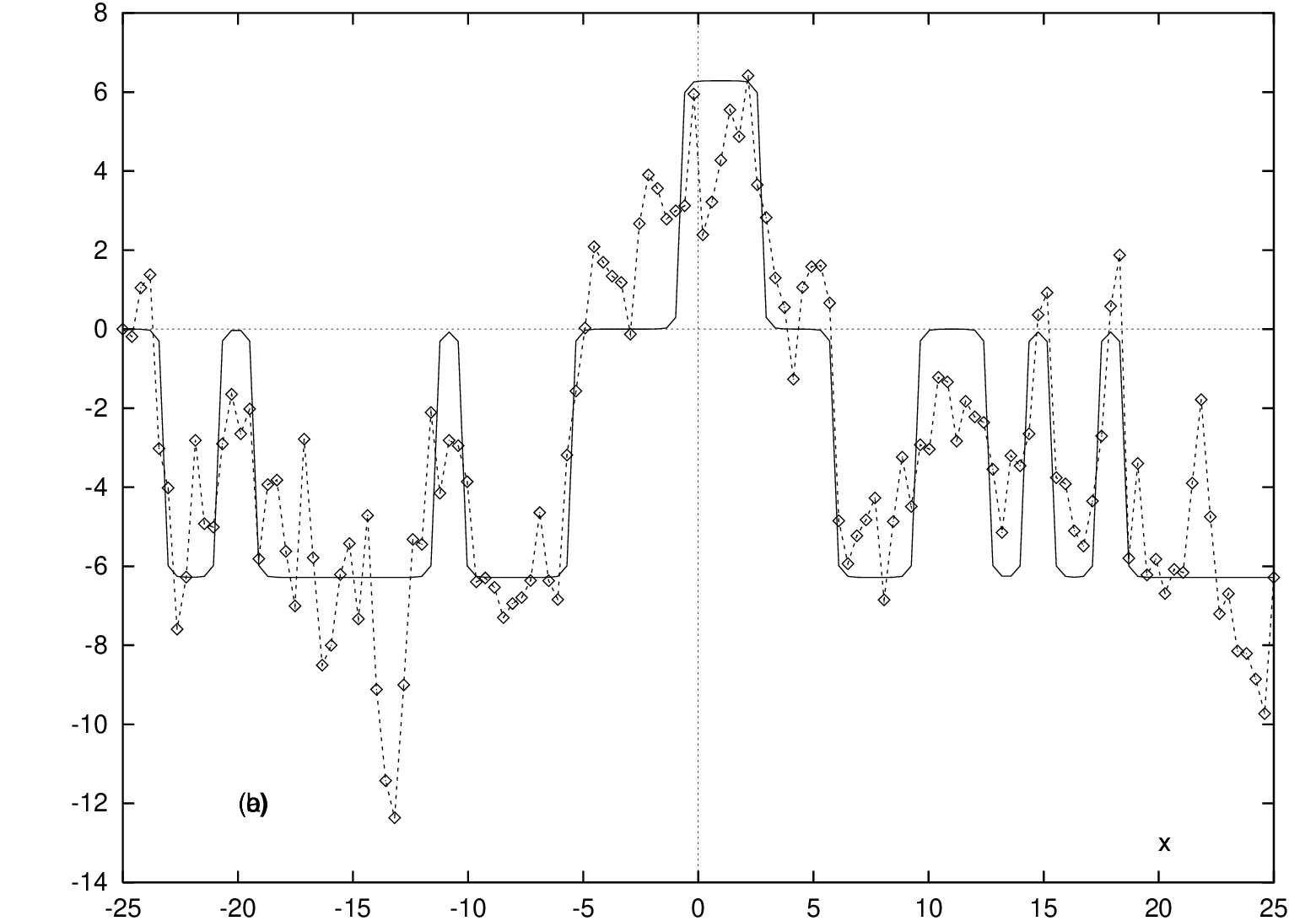}}
\end{center}
\caption{The angular field $\phi(x,t)$ at time $t=20$ (full lines) 
after evolving as in fig.5 in $S^1 \times R^1$ embedding of
the $O(2)$ model from the same random initial configuration $\phi(x,t=0)$ 
(linespoints) into 
quasistable soliton-antisoliton configurations $\phi(x,t=20)$
for increasing symmetry breaking: ~(a): $m_\pi=100$ MeV, 
~(b): $m_\pi=200$ MeV, ~(c): $m_\pi=600$ MeV.}
\label{fig7} 
\end{figure}

Fig.5 shows for the angular evolution the typical situation ($t=20$)
after the second phase: The rapid fluctuations of the ($t=0$) random walk 
in the angle $\phi(x,t)$ (linespoints) are smoothed away, 
the resulting smooth curve (full line) still
follows the average path of the random walk, necessarily always
connecting the fixed boundary values and thus conserving the 
winding number ($B=-1$ in our example). The corresponding trajectories
at the same point in time ($t=20$) for both evolutions are compared 
in figs.6a,b. The radius of both configurations has settled already
at the final limit $r=1$ (in units of $f_\pi$), the angle of the
angular evolution (a) (as shown in fig.5) curls several times back and
forth around the origin with net winding $-2\pi$. In contrast to this
the trajectory (b) of the cartesian evolution has at an early stage of
the evolution suffered several changes of winding number and is
approaching a final configuration with $B=+1$.

As discussed in the previous section, in this simple 1+1 dimensional
model we have to break the $O(2)$ symmetry explicitly in order to
destabilize the static solutions and allow for varying angular winding
length $L_w$ of the static solitons. 
From (\ref{ineq}) the critical destabilizing value is
$m_\pi=m_\sigma/6.5$ MeV, so for values $m_\pi>154$ MeV
the cartesian evolution rapidly ends at the trivial solution
$\sigma=1,~~\pi=0$ with $B=0$. For $m_\pi<154$ MeV there is a
chance for nonvanishing final values of $B$, but they are less likely
because already in the very early stage of the cartesian evolution
the broken symmetry drags the whole initial configuration across
the origin towards the minimum of the potential. 

\begin{figure}[h]
\begin{center}
\leavevmode
\vbox{\epsfxsize=13truecm
\epsfysize=6truecm \epsfbox{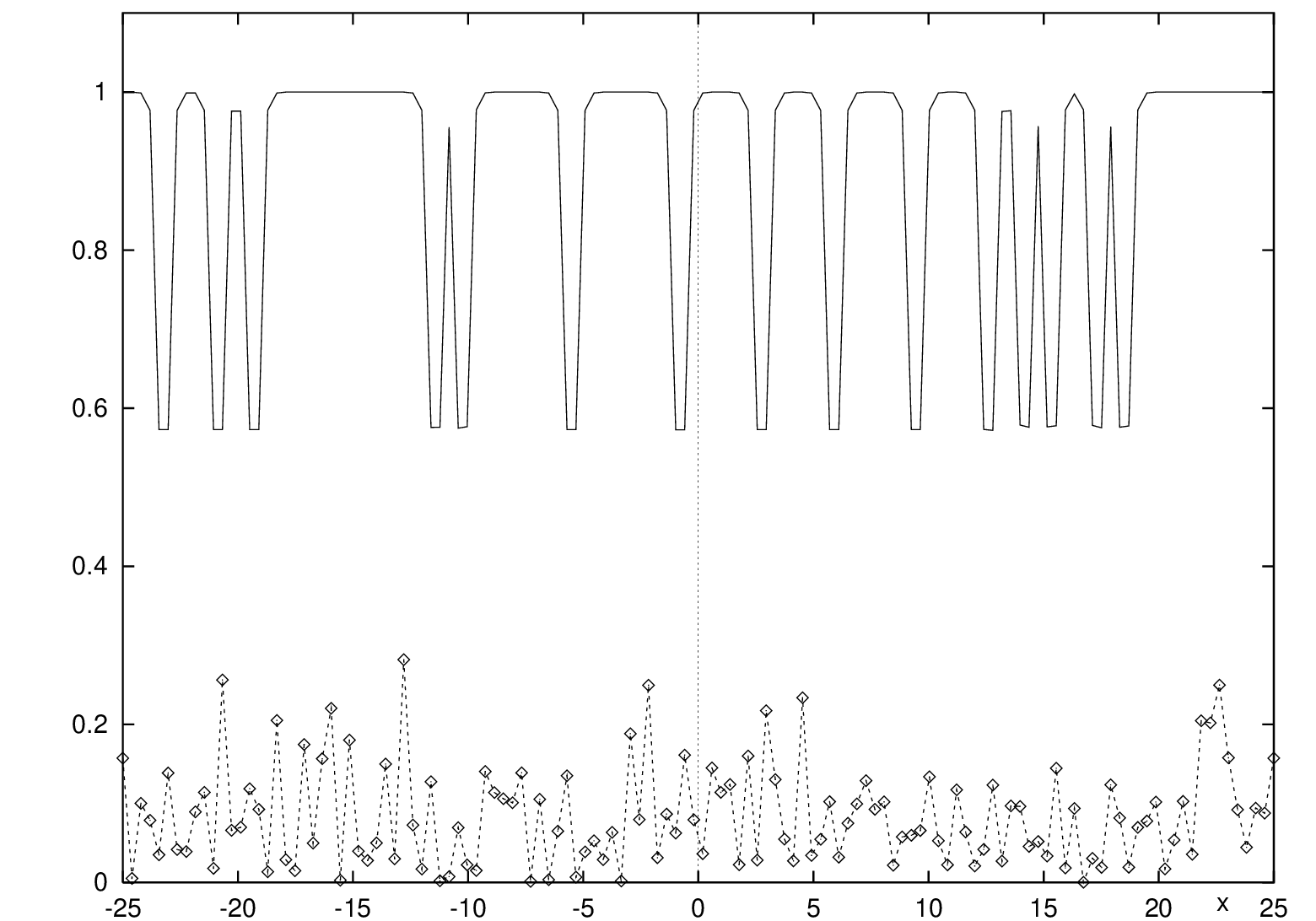}}
\end{center}
\caption{The radial field $r(x,t)$ (in units of $f_\pi$) at time $t=20$ 
after evolving in $S^1 \times R^1$ embedding of
the $O(2)$ model from the random initial configuration $\phi(x,t=0)$ 
(linespoints) as in figs.6 into the 
quasistable soliton-antisoliton configurations $r(x,t=20)$ (full
line) for strong symmetry breaking $m_\pi=600$ MeV, 
corresponding to fig.7c. }
\label{fig8} 
\end{figure}

In contrast to this,
in the angular evolution $B$ is conserved for all values of $m_\pi$ 
and the decreasing angular winding length of the soliton solutions
allows the evolution to approximate the original random fluctuations 
by an increasing number of metastable kink and antikink configurations.
To demonstrate this effect we
compare in figs.7a,b,c  the angular field $\phi(x,t)$ resulting from
the angular evolution at time $t=20$ (as in fig.5) for three values of
the symmetry-breaking 'pion' mass $m_\pi= 100,200,600$ MeV. 
For $m_\pi=100$ MeV the angular field at time $t=20$ has evolved into a 
$n_+=1,~n_-=2$ kink-antikink configuration shown in fig.7a
which firmly rests in a local minimum of the energy hypersurface 
such that it does not evolve any further.

For $m_\pi$-values above 154 MeV we ideally expect a
sequence of step functions in the angular distribution, while in the
radial field the solitons still should have finite spatial extent 
(cf.(\ref{radeq})). This is shown in figs.7b,c for the angular evolution
at time $t=20$ with $m_\pi=200$ and $600$ MeV. 
Again the resulting configurations rest quasistable in local energy minima.
In the numerical evolution due to 
the finiteness of the grid the kinks in the angle cannot really 
become step functions and
the radial bags do not reach the value of $r=0$ at their center (cf. fig.8).
It is natural to expect that also in the continuum limit, when the 
random walk initial configurations approach continuous differentiable
functions the evolution will lead to trajectories which approximate 
these functions by a series of kinks and antikinks, or step functions
(if the symmetry breaking exceeds the critical value (\ref{ineq})),
with conserved net winding. Their local density will depend on the
ratio of their winding length to the size of the box, and on the 
local gradients of the initial trajectory, i.e. on the initial local 
'baryon' density.

\section{Conclusion}
In the commonly used $R^4$ embedding of the linear $O(4)$ $\sigma$ 
model the angular nature of the chiral field is lost.
As a consequence
the winding number is not conserved and therefore no longer can be 
identified with baryon number, the latter concept being 
well established in the $SU(2) \times SU(2)$ nonlinear $\sigma$ model. 

For the early stages in the evolution of a hot hadron gas after a quench
from the symmetric phase into the spontaneously symmetry-breaking cold phase
this can be of crucial importance, because
structures with nontrivial winding number contained in the
random initial configurations can trivially unwind 
in the $R^4$ embedding and thus allow for
the formation of large domains with uniform orientation of the 
chiral field. This has led to suggest spontaneous formation of 
macroscopic domains with differently oriented chiral condensates
if the explicit symmetry breaking is sufficiently small.  

On the other hand, in $S^3 \times R^1$ embedding,
trajectories which are constrained by fixed
boundary conditions on the angular variables cannot unwind and thus 
always reflect the net baryon number of the considered domain.
Dissipative terms which must be present in the evolution equations
will produce smooth trajectories which approximate the initial
random angular configurations by random sequences of stable soliton and 
antisoliton structures. Their density depends crucially on the ratio
of their angular winding length $L_w$ to the size of the domain
considered, but for sufficiently small $L_w$ they closely reflect
the local winding structure of the random initial configuration, as in
the Kibble mechanism.

In the 1+1 dimensional $O(2)$ model used here for illustration the 
angular winding length of nontrivial soliton solutions
is tied to explicit symmetry breaking. For $m_\pi=0$ 
the final trajectories evolving after a quench interpolate
linearly between the values $\phi(-L)$ and $\phi(L)$ at the 
boundaries of the domain with size $2L$.
With explicit symmetry breaking these differ by multiples of
$2\pi$ for sufficiently large domains. 
Small angular winding lengths $L_w$ require large symmetry breaking.
So the evolution of random initial configurations into
soliton- antisoliton sequences within a given domain can be studied,  
but disoriented domains cannot be obtained within this simple model.
This deficiency is absent in 3+1 dimensional $O(4)$ models where 
the mechanism which determines the angular winding length
of nontrivial structures is largely independent of
explicit symmetry breaking.

To summarize, if we want to maintain the topological concept of 
baryon number conservation in 
effective chiral theories near the chiral phase transition, 
we have to observe the angular nature of the
chiral field and allow for a multitude of distinct sets of angles which
describe the same physical situation.
Partition functions in the usual $R^4$ embedding of the $O(4)$ 
model contain trajectories with fluctuating values of baryon number $B$
with comparable weight, while the angular embedding allows to
select also near and above $T_c$ only configurations with the same 
baryon number that characterizes the system at $T=0$.
This feature is a natural ingredient of the nonlinear $\sigma$ model 
for pions and it should not be lost when the chiral partner of the
pion is included as important degree of freedom near $T_c$.

\acknowledgements
The author would like to thank H. Walliser, H. B. Geyer and
F. Meier for many helpful discussions.

\end{document}